\documentclass{article}

\usepackage{microtype}
\usepackage{graphicx}
\usepackage{subcaption}
\usepackage{booktabs} 

\usepackage{hyperref}

\usepackage[preprint]{icml2026}

\usepackage{amsmath}
\usepackage{amssymb}
\usepackage{mathtools}
\usepackage{amsthm}
\usepackage{multirow}

\usepackage[capitalize,noabbrev]{cleveref}

\theoremstyle{plain}

\theoremstyle{definition}

\theoremstyle{remark}

\usepackage[textsize=tiny]{todonotes}

\newcommand{\modelname}[1]{{\sc FCDM}{#1}}

\icmltitlerunning{\modelname: A Physics-Guided Bidirectional \underline{F}requency Aware \underline{C}onvolution and \underline{D}iffusion-Based \underline{M}odel for Sinogram Inpainting}

\begin{document}

\twocolumn[
  \icmltitle{\modelname: A Physics-Guided Bidirectional \underline{F}requency Aware \underline{C}onvolution and \underline{D}iffusion-Based \underline{M}odel for Sinogram Inpainting}



  \icmlsetsymbol{equal}{*}

  \begin{icmlauthorlist}
    \icmlauthor{Jiaze E}{wm}
    \icmlauthor{Srutarshi Banerjee}{anl}
    \icmlauthor{Tekin Bicer}{anl}
    \icmlauthor{Guannan Wang}{wm}
    \icmlauthor{Yanfu Zhang}{wm}
    \icmlauthor{Bin Ren}{wm}
  \end{icmlauthorlist}

  \icmlaffiliation{wm}{William \& Mary, USA}
  \icmlaffiliation{anl}{Argonne National Laboratory, USA}

  \icmlcorrespondingauthor{Bin Ren}{bren@wm.edu}

  \icmlkeywords{Machine Learning, ICML}

  \vskip 0.3in
]



\printAffiliationsAndNotice{}  

\begin{abstract}
Computed tomography (CT) is widely used in scientific imaging systems such as synchrotron and laboratory-based nano-CT, but acquiring full-view sinograms requires high radiation dose and long scan times. Sparse-view CT reduces this burden but produces incomplete sinograms with structured signal loss, degrading reconstruction quality. Unlike RGB images, sinograms encode globally coupled projections and exhibit directional spectral patterns, making conventional RGB-oriented inpainting methods—including diffusion models—ineffective, as they ignore angular dependencies and physical constraints inherent to tomographic data. We propose~\modelname, a diffusion-based framework for sinogram restoration that incorporates bidirectional frequency reasoning, angular-aware masking, and physics-guided regularization to preserve global structure and physical plausibility. Experiments on real-world datasets show that~\modelname~consistently outperforms existing baselines, achieving over 0.93 SSIM and 31 dB PSNR across diverse sparse-view settings.
\end{abstract}

\section{Introduction}

X-ray computed tomography (XCT) is a vital 3D imaging technique widely used in synchrotron and laboratory-based nano-CT systems for high-resolution internal structure analysis. It enables studies of battery degradation \cite{zhao2024suppressing,muller2018quantification}, neurodegenerative disease studies \cite{koroshetz2018state,dyer2017quantifying,hidayetouglu2020petascale}, and materials characterization \cite{satapath2024porosity}. Synchrotron XCT uses a parallel-beam geometry, ensuring quantitative accuracy and uniform illumination. The projections form sinograms, reconstructed into 3D volumes through computational algorithms \cite{chen2019ifdk,hidayetouglu2019memxct,bicer2017trace}. However, acquiring full-view projections requires long scan times and high radiation doses \cite{brenner2007computed}, which can degrade samples and limit throughput. Sparse-view CT alleviates this burden by reducing projections, but incomplete sinograms introduce artifacts and structural loss in reconstruction. Neither increasing projection density—often infeasible for fragile samples—nor post-reconstruction correction can recover the lost information, since missing sinogram data cause nonlinear distortions that amplify during reconstruction. Restoring missing data directly in the projection domain, i.e., sinogram inpainting, is therefore essential for accurate reconstruction.

\begin{figure}[t]
    \centering
    \begin{subfigure}{0.1\textwidth}
        \centering
        \includegraphics[width=\textwidth]{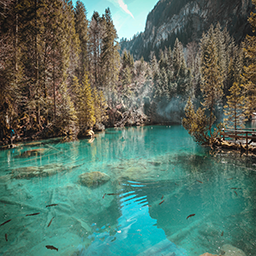}
        \caption{RGB Img}
    \end{subfigure}
    \hfill
    \begin{subfigure}{0.1\textwidth}
        \centering
        \includegraphics[width=\textwidth]{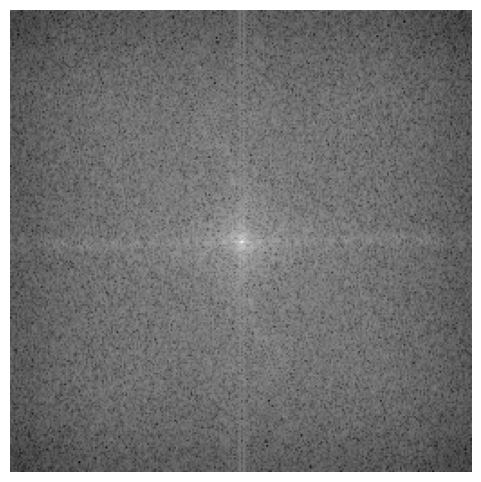}
        \caption{RGB Spec}
    \end{subfigure}
    \hfill
    \begin{subfigure}{0.1\textwidth}
        \centering
        \includegraphics[width=\textwidth]{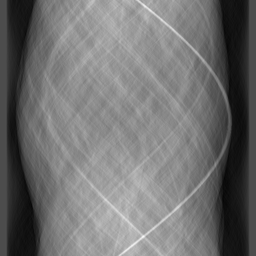}
        \caption{Sino Img}
    \end{subfigure}
    \hfill
    \begin{subfigure}{0.1\textwidth}
        \centering
        \includegraphics[width=\textwidth]{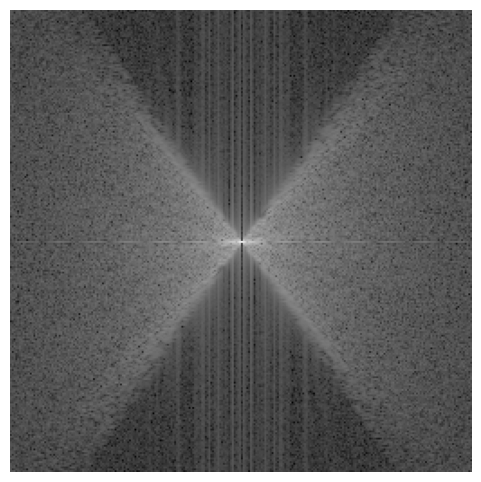}
        \caption{Sino Spec}
    \end{subfigure}
    \caption{Comparisons of RGB, sinogram and their spectra. Unlike RGB images that have localized frequency components, sinograms exhibit structured spectral distributions due to the Radon transform.}
    \label{fig:spec}
\end{figure}

Although inpainting methods have been extensively studied in the context of RGB images \cite{liu2024structure, zhang2023coherent, ko2023continuously, deng2022hourglass, lugmayr2022repaint, li2022misf, suvorov2022resolution}, their applicability to X-ray sinograms remains relatively underexplored. Most RGB-based approaches rely on the assumption that missing regions can be locally inferred from surrounding pixels. However, this assumption does not readily hold for sinograms, where each pixel encodes an integrated projection value along an X-ray path, leading to globally coupled and highly entangled structures. Several studies have explored sinogram completion using U-Net-based \cite{zhao2018unsupervised, yao2024no}, GAN-based \cite{ valat2023sinogram, xie2022limited} or Transformer-based \cite{jiaze2025sinotx} models, and some have incorporated periodicity or reconstruction-based constraints \cite{li2019sinogram, wagner2023geometric}. However, these methods remain largely spatial, focusing on pixel-wise interpolation or appearance realism, and fail to explicitly model the structured frequency domain or the governing physics of sinogram formation, which are fundamentally distinct from those of RGB images. As defined by the Radon transform \cite{radon_uber_1917}, the detector and angle axes of a sinogram represent distinct physical dimensions, leading to highly directional and asymmetric spectral patterns (see Figure~\ref{fig:spec}). These observations highlight the need to explicitly account for the frequency structure and physical consistency inherent to sinograms when performing data completion.

Building on the success of diffusion models in image synthesis and restoration \cite{ho2020denoising, lugmayr2022repaint, rombach2022high}, we integrate them into our framework for sinogram completion to leverage their strong capability in modeling complex data distributions and generating globally coherent structures through iterative denoising. However, direct application is hindered by the modality gap: in RGB inpainting, missing pixels are local and independent, while sinogram measurements are globally coupled across projection angles. Moreover, CT reconstruction is more sensitive to low-frequency perturbations due to the integral nature of the inverse Radon transform \cite{wurfl2016deep}. These factors make standard diffusion models—built for spatially independent and uniformly noised data—less effective for sinogram restoration, motivating a geometry- and frequency-aware design.

To address these limitations, we develop~\modelname, a diffusion-based framework for sinogram inpainting that incorporates domain-specific frequency organization and physical constraints. Rather than introducing isolated heuristics,~\modelname~systematically addresses the failure of standard diffusion at three coupled levels: representation learning, conditional guidance, and noise injection. At the representation level,~\modelname~learns a physically consistent latent representation prior to denoising, avoiding direct diffusion in the high-dimensional projection space where low-frequency distortions can be amplified. To encode the anisotropic frequency structure induced by the Radon transform, we employ a bidirectional frequency-domain convolution that processes detector- and angle-wise features independently. Physical validity is further encouraged through lightweight physics- and frequency-based regularization, including a total-absorption conservation constraint \cite{kak2001principles}. Together, these components provide a suitable representation for diffusion-based restoration.

At the denoising stage,~\modelname~applies diffusion to refine the inpainting; however, under sinogram geometry, the standard assumption of spatial independence no longer holds, as missing regions span correlated projection angles. To address this issue, we enhance conditional guidance by encoding projection-angle information into the mask representation, enabling the denoising process to reason about missing data across views and align incomplete projections with observed structures. In parallel, the noise injection process is adapted through frequency-aware scheduling, which preserves low-frequency components at early steps to stabilize global recovery and progressively emphasizes high-frequency refinement. Together, these designs make the diffusion process both angularly aware and frequency adaptive, yielding restorations that remain consistent across projection angles and robust to spectral perturbations.

In summary, our main contributions are as follows:
\begin{itemize}
    \item We propose~\modelname, a diffusion-based framework that jointly incorporates sinogram frequency organization and physical constraints, addressing feature entanglement and physical inconsistency overlooked by prior inpainting methods.
    \item We introduce a bidirectional frequency-domain convolution to disentangle spectral features along detector and angular dimensions, together with a physics-guided loss enforcing total absorption conservation \cite{kak2001principles}.
    \item We further align diffusion with sinogram geometry through Fourier-enhanced mask embedding that encodes angular context into missing regions, and frequency-adaptive noise scheduling that balances global structure recovery and high-frequency refinement.
    \item Extensive experiments on TomoBank and LoDoPaB \cite{de2018tomobank,leuschner2021lodopab} show that~\modelname~consistently outperforms existing baselines, achieving over 0.93 SSIM and 31 dB in PSNR, with ablations validating each component.
\end{itemize}

\section{Related Work}

Deep learning-based methods have shown great promise in RGB image inpainting. \cite{zhang2023coherent} proposes a joint optimization method based on a Bayesian framework. \cite{ko2023continuously} applies a continuous mask on Transformer-based structure. \cite{deng2022hourglass} proposes a new Hourglass attention structure and combines it with the Transformer structure. \cite{li2022misf} presents a new inpainting framework with multi-level interactive siamese filtering using an interactive dynamic kernel prediction mechanism. \cite{suvorov2022resolution} proposes LaMa which utilizes fast Fourier convolution \cite{NEURIPS2020_2fd5d41e} to solve large masking problmes. \cite{lugmayr2022repaint} proposes an inpainting method that solely leverages an off-the-shelf unconditionally trained denoising diffusion probabilistic models (DDPM). \cite{liu2024structure} uses diffusion-based structure guidance to solve the problem of semantic discrepancy in masked and unmasked regions. However, RGB image inpainting methods do not consider the unique physical constraints and feature entanglements in sinograms.

Traditional mathematical methods have also been employed for sinogram inpainting. Linear \cite{herman1993algebraic} and spline interpolation \cite{unser1999splines} estimate the missing sinogram data based on the known data. Total Variation (TV) minimization \cite{chambolle1997image}, originally developed for image denoising and reconstruction, has also been applied to sinogram inpainting. It works by promoting sparsity in the gradient domain, making it effective for inpainting sinograms with small missing regions. While these mathematical methods provide simple and interpretable solutions, they struggle with large missing regions.

Recently, several deep learning-based models have been proposed for sinogram inpainting. \cite{yao2024no, zhao2018unsupervised} utilize U-Net to directly interpolate missing sinogram data. \cite{tomography9030094, s19183941, valat2023sinogram, xie2022limited} employ GANs to generate realistic sinogram patches. \cite{jiaze2025sinotx} uses Transformer blocks to inpaint geometric shapes sinograms. However, these methods typically operate in the spatial domain and may struggle to enforce global consistency or preserve underlying frequency characteristics. Other deep learning-based models apply unique characteristics of sinograms. \cite{li2019promising, li2019sinogram} incorporate periodicity into the loss function to leverage the cyclical nature of certain sinograms. \cite{wagner2023geometric, s19183941} compare the reconstructed images from the inpainted and ground truth sinograms in their loss calculations. \cite{9606601} combines coordinate mapping and Fourier feature mapping together with multilayer perceptron to help the inpainting. While these strategies improve inpainting fidelity to some extent, they generally lack explicit modeling of sinogram behavior in the frequency domain and do not directly regulate the progression of the inpainting process. Several recent approaches take a different direction by bypassing sinogram inpainting entirely and instead reconstructing images directly from incomplete sinograms using deep generative models. Representative works include \cite{liu2023dolce, hu2023cross, guo2025advancing}, which refine reconstructed images through learned priors or image-space denoising. While effective in suppressing reconstruction artifacts, these end-to-end models directly map incomplete sinograms to reconstructed images, thereby bypassing explicit sinogram recovery and losing direct supervision or control over projection-domain consistency.

\section{FCDM Design and Analysis}

\begin{figure}[t]
    \centering
    \begin{subfigure}{0.45\textwidth}
        \centering
        \includegraphics[width=\textwidth]{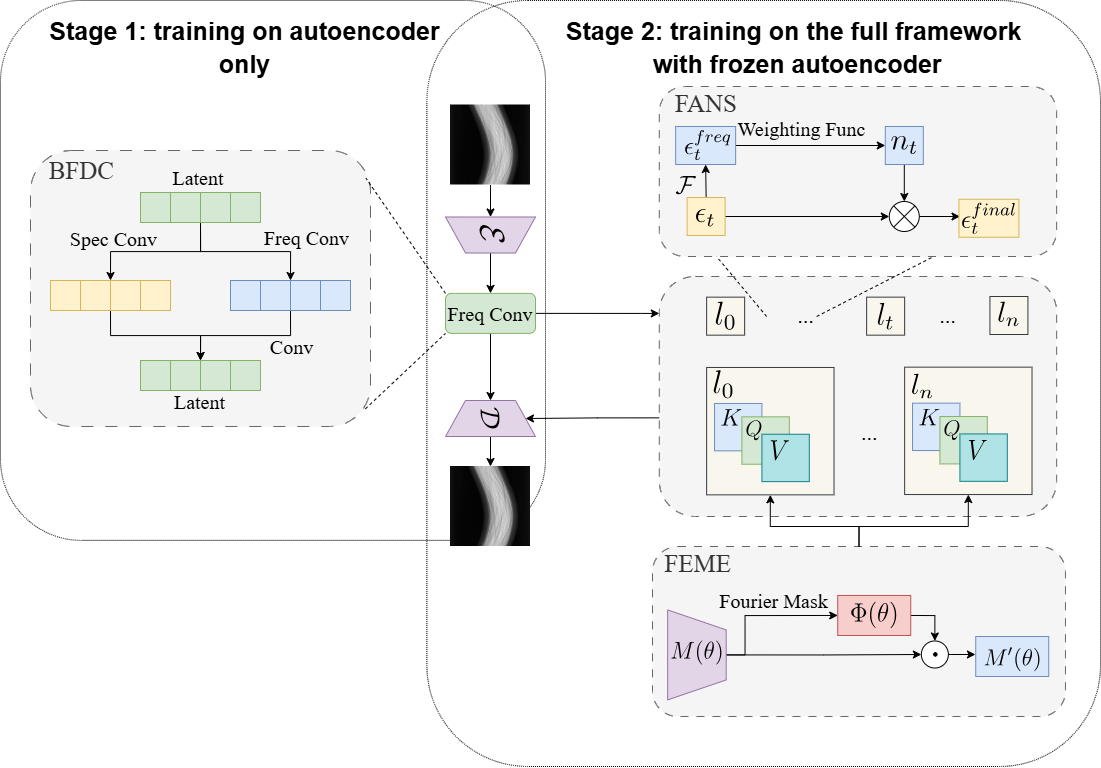}
    \end{subfigure}
    \caption{Overview of~\modelname. Stage 1 trains an encoder–decoder network equipped with BFDC to extract frequency– and geometry–aware latent representations. Stage 2 applies a diffusion-based model to perform latent-space inpainting guided by FEME and FANS. Here, $\mathcal{E}$ / $\mathcal{D}$ denote the encoder / decoder. $l_0$, $l_t$, $l_N$ denote latents in the diffusion process. $K$, $Q$, $V$ are query, key, and value matrices from the attention mechanism. $\mathcal{F}$ denote the Fourier transforms. $n_t$ is the time-dependent spectral weighting map. $\epsilon_t$, $\epsilon_t^{freq}$, and $\epsilon_t^{final}$ are the standard, frequency-adapted and final noises. $M(\theta)$ is the angle-dependent input mask, $\Phi(\theta)$ is the Fourier encoding operator applied to it, and $M'(\theta)$ is the resulting Fourier-enhanced mask embedding produced by FEME.}
    \label{fig:overview}
\end{figure}

\modelname~is a two-stage framework for sinogram inpainting, combining an encoder–decoder network with a diffusion-based restoration process. As shown in Figure~\ref{fig:overview}, stage 1 focuses on representation learning, and stage 2 performs guided denoising for inpainting. In stage 1, the framework learns a latent representation of the input sinogram. The encoder maps the sinogram into a latent space, which is processed by bidirectional frequency-domain convolutions (BFDC) to extract complementary spatial and frequency-domain features across projection angles and detector positions. These enriched features are then decoded to recover the sinogram. The encoder–decoder is trained using pixel-wise and adversarial losses, augmented by two additional terms that promote frequency consistency and physical realism. After training, the autoencoder parameters are frozen to provide stable feature extraction for the stage 2. In stage 2, a diffusion-based model is trained to perform inpainting within the learned latent space. During training, random binary masks simulate missing or corrupted sinogram regions with various angular spans. The mask information is injected into the model through a Transformer-style attention mechanism to guide conditional denoising. To improve this process, we introduce Fourier-enhanced mask embedding (FEME), which encodes the projection-angle information of missing regions into the mask representation, enhancing geometric awareness. We further design Frequency-adaptive noise scheduling (FANS), which adjusts the diffusion noise by frequency sensitivity—preserving low-frequency components during early denoising for structural stability and gradually emphasizing high-frequency details to refine fine textures. Together, the two stages establish a unified pipeline that first learns frequency– and physics–aware representations, then performs geometry-consistent inpainting within the latent space for accurate and physically coherent sinogram restoration.

\subsection{Bidirectional Frequency-Domain Convolutions}

To enhance feature extraction in the latent space, we introduce bidirectional frequency-domain convolutions (BFDC). Given a sinogram input $s \in \mathbb{R}^{C \times H \times W}$, the encoder $\mathcal{E}(\cdot)$ maps it to a latent representation $h = \mathcal{E}(s), h \in \mathbb{R}^{C' \times H' \times W'}$. We then apply both spatial and frequency-domain convolutions to $h$. A standard spatial convolution $\mathcal{G}(\cdot)$ produces a spatial representation $h_s = \mathcal{G}(h)$. We also perform frequency-domain convolutions along two orthogonal dimensions—detector ($H^{\prime}$) and projection angle ($W^{\prime}$)—-to capture directional spectral features that are entangled in the sinogram.

Specifically, we apply real-valued Fourier transforms (RFFT) \cite{cooley1965algorithm} along the width and height axes of $h$, yielding complex-valued representations $\hat{h}_w$ and $\hat{h}_h$. These are multiplied with corresponding frequency-domain kernels $\mathcal{K}_w$ and $\mathcal{K}_h$ via Hadamard product, then transformed back to the spatial domain using an inverse RFFT to obtain frequency-enhanced features:
\begin{equation}
    l_w = \mathcal{F}_w^{-1}(\mathcal{K}_w \cdot \mathcal{F}_w(h)), \quad l_h = \mathcal{F}_h^{-1}(\mathcal{K}_h \cdot \mathcal{F}_h(h)).
\end{equation}

Unlike full 2D Fourier convolutions, which treat the two axes equally and may obscure their distinct physical meanings, our directional decomposition allows independent processing of features along projection and detector axes. The final representation is computed as a weighted fusion of the spatial and frequency-enhanced components:
\begin{equation}
    l = h_s + \alpha_w l_w + \alpha_h l_h,
\end{equation}
where $\alpha_w$ and $\alpha_h$ are weighting factors that ensure frequency-domain information along both projection angles and detector positions are optimally integrated. The fused representation $l$ is then decoded by the decoder $\mathcal{D}(\cdot)$ to produce the reconstructed sinogram $\overline{s} = \mathcal{D}(l)$.

\subsection{Physics-Guided Loss Functions}

\subsubsection{Total Projection Consistency Loss}

In CT, the total X-ray absorption along each projection should remain consistent with the object's integral attenuation. This ensures that no non-physical intensity drift is introduced, which could distort the sinogram inpainting and further the final CT reconstruction. To enforce this constraint, we require for each angle, the sum of all projection values to match the total absorption of the projection data:
\begin{equation}
    \forall \theta \int P_\theta(r)\, dr = \int \mu(x,y)\, dx\, dy.
\end{equation}
Here, $P_{\theta}(r)$ is the projection at angle $\theta$ and detector position $r$, and $\mu(x,y)$ is the object's attenuation coefficient. Since $\mu(x,y)$ is not directly available during training, we approximate the total absorption via the inverse Radon transform $\mathcal{R}^{-1} \{\bar{P}\}$, and define the following loss:
\begin{equation}
    \mathcal{L}_{absorp} = \sum_{\theta} \left( \int \bar{P}_{\theta}(r)\, dr - \int \mathcal{R}^{-1}\{\bar{P}\}(x,y)\, dx\, dy \right)^2 .
\end{equation}
This term penalizes global mismatches between the measured projections and their corresponding inferred attenuation distribution, encouraging physically consistent inpainting.

\subsubsection{Frequency Domain Consistency Loss}

Since sinogram information is naturally structured in the frequency domain, ensuring spectral consistency between inpainted and ground-truth sinograms is essential. We define this constraint by minimizing the L2 norm of the frequency difference between the inpainted and ground-truth sinograms:
\begin{equation}
    \mathcal{L}_{\text{freq}} = \sum_{u} \sum_{v} \left\{ \mathcal{F}(\overline{\mu}(x,y)) - \mathcal{F}({\mu}(x,y)) \right\}^2.
\end{equation}
where $\mathcal{F}$ denotes the 2D Fourier transform. This loss ensures that the framework accurately captures the structural characteristics of sinograms in the frequency domain, preventing spectral artifacts.

\subsubsection{Overall Loss}

The final loss for stage 1 training combines pixel-wise reconstruction, adversarial learning, and physics-guided constraints:
\begin{equation}
    \mathcal{L}_1 = \lambda_{\text{pixel}} \mathcal{L}_{\text{pixel}} + \lambda_{\text{adv}} \mathcal{L}_{\text{adv}} + \lambda_{\text{absorp}} \mathcal{L}_{\text{absorp}} + \lambda_{\text{freq}} \mathcal{L}_{\text{freq}}.
\end{equation}
where $\mathcal{L}_{\text{pixel}}$ refers to mean squared error (MSE) loss that ensures pixel-wise reconstruction accuracy, and $\mathcal{L}_{\text{adv}}$ is the adversarial loss. In stage 2, the diffusion model optimizes a MSE loss.

\subsection{Fourier-Enhanced Mask Embedding}

Standard diffusion-based inpainting methods use binary 0/1 masks to indicate missing regions. Unlike RGB images, sinograms exhibit global feature entanglement due to the projection process. The same physical structure may appear at different spatial positions across projection angles, while a simple binary mask offers no indication of these angular correlations. As a result, the framework may interpret missing areas as independent holes, failing to capture their underlying coherence. To address this, we introduce Fourier-enhanced mask embedding (FEME), which augments each mask entry with direction-aware features to better reflect the sinogram’s projection geometry. Instead of treating projection angles as raw scalar inputs, we encode each angle using a Fourier basis, capturing the periodic nature of angular variation. Given a projection angle $\theta$, we compute its Fourier representation as  
\begin{equation}
    \Phi(\theta) = \left[\sin(2\pi f_1 \theta), \cos(2\pi f_1 \theta), \dots, \cos(2\pi f_N \theta) \right],
\end{equation}
where $f_i$ represents a set of predefined frequencies. This encoding transforms scalar angles into a structured representation that preserves their cyclical relationships and enhances expressiveness. The encoded angle is broadcast to each detector position and concatenated with the binary mask $M(\theta)$, which is expanded from a scalar to a vector to match the dimension:
\begin{equation}
    M'(\theta) = [M(\theta) ; \Phi(\theta)].
\end{equation}

This operation enriches the original binary mask with projection-specific context. By embedding angular information in a periodic, high-dimensional form, FEME enables the framework to better identify structural correspondences across projections, leading to more coherent and physically consistent inpainting results.

\subsection{Frequency-Adaptive Noise Scheduling}

\begin{table*}[t]
    \centering
    \caption{Quantitative comparison of SSIM and PSNR on {\tt TomoBank} and {\tt LoDoPaB} under random and periodic masking with ratios of 0.4, 0.6, and 0.8. Each entry reports the scores of the inpainted sinogram followed by its FBP \cite{ramachandran1971three} reconstruction in parentheses.}
    \label{tab:comparison}
    \scriptsize
    \begin{tabular}{llccccccc}
        \hline
        \multirow{2}{*}{Method} & \multirow{2}{*}{} & \multirow{2}{*}{Mask} &
        \multicolumn{3}{c}{SSIM} &
        \multicolumn{3}{c}{PSNR} \\
        \cline{4-9} & & & 0.4 & 0.6 & 0.8 & 0.4 & 0.6 & 0.8 \\
        \hline
        \multicolumn{8}{l}{TomoBank}\\
        \hline
        \multirow{2}{*}{Ours} & \multirow{2}{*}{} & Random & \textbf{0.940} (0.919) & \textbf{0.938} (0.918) & \textbf{0.934} (0.915) &
        \textbf{31.3} (29.4) & \textbf{31.1} (29.2) & \textbf{30.7} (29.1) \\
        & & Periodic & 0.937 (0.916) & 0.935 (0.915) & 0.933 (0.913) &
        31.0 (29.1) & 30.8 (29.0) & 30.6 (28.8) \\
        \multirow{2}{*}{RePaint} & \multirow{2}{*}{CVPR 2022} &  Random & 0.928 (0.906) & 0.925 (0.905) & 0.923 (0.902) &
        29.9 (27.5) & 29.8 (27.4) & 29.5 (27.2) \\
        & & Periodic & 0.926 (0.904) & 0.924 (0.902) & 0.921 (0.901) &
        29.8 (27.3) & 29.6 (27.1) & 29.4 (27.1) \\
        \multirow{2}{*}{CoPaint} & \multirow{2}{*}{ICML 2023} & Random & 0.918 (0.896) & 0.915 (0.893) & 0.912 (0.890) &
        29.2 (27.3) & 29.0 (27.2) & 28.8 (27.0) \\
        & & Periodic & 0.915 (0.893) & 0.913 (0.891) & 0.910 (0.889) &
        29.0 (27.1) & 28.8 (27.0) & 28.7 (26.8) \\
        \multirow{2}{*}{StrDiffusion} & \multirow{2}{*}{CVPR 2024} & Random & 0.911 (0.888) & 0.908 (0.885) & 0.905 (0.883) &
        29.0 (27.1) & 28.9 (26.9) & 28.6 (26.7) \\
        & & Periodic & 0.909 (0.886) & 0.907 (0.884) & 0.904 (0.882) &
        28.8 (27.0) & 28.7 (26.8) & 28.5 (26.6) \\
        \multirow{2}{*}{MISF} & \multirow{2}{*}{CVPR 2022} & Random & 0.905 (0.883) & 0.902 (0.880) & 0.899 (0.878) &
        29.1 (27.2) & 28.8 (27.0) & 28.5 (26.8) \\
        & & Periodic & 0.904 (0.881) & 0.900 (0.879) & 0.897 (0.877) &
        28.9 (27.1) & 28.7 (26.9) & 28.5 (26.7) \\
        \multirow{2}{*}{LaMa} & \multirow{2}{*}{WACV 2022} & Random & 0.896 (0.872) & 0.893 (0.869) & 0.888 (0.866) &
        28.6 (26.8) & 28.5 (26.7) & 28.3 (26.6) \\
        & & Periodic & 0.895 (0.871) & 0.891 (0.869) & 0.887 (0.866) &
        28.5 (26.8) & 28.4 (26.7) & 28.1 (26.5) \\
        \multirow{2}{*}{SinoTx} & \multirow{2}{*}{ICIP 2025} & Random & 0.829 (0.789) & 0.826 (0.788) & 0.821 (0.786) &
        27.1 (24.4) & 26.8 (24.3) & 26.5 (24.2) \\
        & & Periodic & 0.827 (0.787) & 0.824 (0.786) & 0.819 (0.784) &
        26.9 (24.3) & 26.6 (24.2) & 26.4 (24.1) \\
        \multirow{2}{*}{Hourglass} & \multirow{2}{*}{ECCV 2022} & Random & 0.791 (0.750) & 0.788 (0.748) & 0.785 (0.746) &
        25.5 (23.1) & 25.4 (23.2) & 25.2 (23.0) \\
        & & Periodic & 0.789 (0.749) & 0.786 (0.747) & 0.782 (0.744) &
        25.3 (23.0) & 25.2 (23.1) & 25.0 (22.9) \\
        \multirow{2}{*}{UsiNet} & \multirow{2}{*}{NPJ 2024} & Random & 0.635 (0.602) & 0.631 (0.599) & 0.628 (0.595) &
        21.9 (19.7) & 21.6 (19.6) & 21.3 (19.3) \\
        & & Periodic & 0.633 (0.600) & 0.630 (0.598) & 0.626 (0.594) &
        21.8 (19.6) & 21.5 (19.5) & 21.2 (19.2) \\
        \multirow{2}{*}{CoIL} & \multirow{2}{*}{IEEE TCI 2021} & Random & 0.741 (0.701) & 0.737 (0.698) & 0.734 (0.695) &
        23.8 (21.5) & 23.6 (21.4) & 23.3 (21.2) \\
        & & Periodic & 0.739 (0.700) & 0.736 (0.697) & 0.732 (0.694) &
        23.7 (21.4) & 23.4 (21.3) & 23.1 (21.1) \\
        \multirow{2}{*}{CMT} & \multirow{2}{*}{ICCV 2023} & Random & 0.716 (0.674) & 0.713 (0.672) & 0.709 (0.669) &
        23.1 (20.6) & 22.9 (20.5) & 22.7 (20.4) \\
        & & Periodic & 0.714 (0.673) & 0.711 (0.670) & 0.707 (0.668) &
        23.0 (20.5) & 22.8 (20.4) & 22.5 (20.3) \\
        \hline
        \multicolumn{8}{l}{Lodopab}\\
        \hline
        \multirow{2}{*}{Ours} & \multirow{2}{*}{} & Random & \textbf{0.947} (0.925) & \textbf{0.944} (0.923) & \textbf{0.940} (0.921) &
        \textbf{31.8} (29.8) & \textbf{31.6} (29.6) & \textbf{31.2} (29.3) \\
        & & Periodic & 0.944 (0.923) & 0.941 (0.921) & 0.939 (0.919) &
        31.6 (29.6) & 31.4 (29.4) & 31.0 (29.2) \\
        \multirow{2}{*}{RePaint} & \multirow{2}{*}{CVPR 2022} & Random & 0.936 (0.912) & 0.933 (0.910) & 0.928 (0.908) &
        30.5 (27.8) & 30.2 (27.7) & 29.8 (27.5) \\
        & & Periodic & 0.934 (0.911) & 0.931 (0.909) & 0.927 (0.907) &
        30.3 (27.6) & 30.0 (27.5) & 29.7 (27.3) \\
        \multirow{2}{*}{CoPaint} & \multirow{2}{*}{ICML 2023} & Random & 0.924 (0.900) & 0.921 (0.897) & 0.918 (0.896) &
        29.7 (27.5) & 29.4 (27.3) & 29.2 (27.1) \\
        & & Periodic & 0.922 (0.899) & 0.919 (0.897) & 0.916 (0.894) &
        29.5 (27.3) & 29.2 (27.2) & 29.0 (27.0) \\
        \multirow{2}{*}{StrDiffusion} & \multirow{2}{*}{CVPR 2024} & Random & 0.916 (0.894) & 0.913 (0.891) & 0.910 (0.888) &
        29.5 (27.3) & 29.2 (27.2) & 29.0 (27.0) \\
        & & Periodic & 0.914 (0.892) & 0.911 (0.889) & 0.908 (0.887) &
        29.3 (27.2) & 29.0 (27.1) & 28.8 (27.0) \\
        \multirow{2}{*}{MISF} & \multirow{2}{*}{CVPR 2022} & Random & 0.909 (0.885) & 0.906 (0.883) & 0.902 (0.880) &
        29.4 (27.2) & 29.1 (27.1) & 28.8 (26.9) \\
        & & Periodic & 0.907 (0.883) & 0.904 (0.881) & 0.901 (0.879) &
        29.2 (27.1) & 28.9 (27.0) & 28.6 (26.8) \\
        \multirow{2}{*}{LaMa} & \multirow{2}{*}{WACV 2022} & Random & 0.899 (0.875) & 0.896 (0.872) & 0.893 (0.869) &
        28.9 (27.0) & 28.7 (26.9) & 28.5 (26.8) \\
        & & Periodic & 0.898 (0.874) & 0.894 (0.871) & 0.891 (0.868) &
        28.7 (26.9) & 28.5 (26.8) & 28.2 (26.6) \\
        \multirow{2}{*}{SinoTx} & \multirow{2}{*}{ICIP 2025} & Random & 0.834 (0.794) & 0.830 (0.791) & 0.827 (0.789) &
        27.4 (24.8) & 27.2 (24.7) & 27.0 (24.5) \\
        & & Periodic & 0.832 (0.792) & 0.828 (0.789) & 0.824 (0.787) &
        27.3 (24.7) & 27.0 (24.6) & 26.8 (24.4) \\
        \multirow{2}{*}{Hourglass} & \multirow{2}{*}{ECCV 2022} & Random & 0.798 (0.756) & 0.795 (0.752) & 0.791 (0.750) &
        25.9 (23.6) & 25.7 (23.4) & 25.4 (23.2) \\
        & & Periodic & 0.796 (0.754) & 0.792 (0.751) & 0.788 (0.748) &
        25.7 (23.5) & 25.5 (23.3) & 25.2 (23.1) \\
        \multirow{2}{*}{UsiNet} & \multirow{2}{*}{NPJ 2024} & Random & 0.641 (0.607) & 0.637 (0.604) & 0.633 (0.600) &
        22.2 (19.9) & 22.0 (19.8) & 21.7 (19.6) \\
        & & Periodic & 0.639 (0.605) & 0.635 (0.602) & 0.631 (0.599) &
        22.1 (19.8) & 21.8 (19.7) & 21.5 (19.5) \\
        \multirow{2}{*}{CoIL} & \multirow{2}{*}{IEEE TCI 2021} & Random & 0.745 (0.705) & 0.742 (0.702) & 0.738 (0.699) &
        24.2 (21.9) & 24.0 (21.8) & 23.7 (21.6) \\
        & & Periodic & 0.743 (0.704) & 0.740 (0.701) & 0.736 (0.698) &
        24.0 (21.8) & 23.8 (21.7) & 23.5 (21.5) \\
        \multirow{2}{*}{CMT} & \multirow{2}{*}{ICCV 2023} & Random & 0.720 (0.678) & 0.717 (0.676) & 0.713 (0.673) &
        23.5 (21.0) & 23.3 (20.9) & 23.1 (20.8) \\
        & & Periodic & 0.718 (0.677) & 0.715 (0.674) & 0.711 (0.671) &
        23.4 (20.9) & 23.2 (20.8) & 22.9 (20.7) \\
        \hline
    \end{tabular}
\end{table*}

In standard diffusion models, Gaussian noise is injected uniformly across all frequency components, ignoring the structural characteristics of the data. Low-frequency components encode global structure, whereas high-frequency components capture fine details, playing distinct roles during denoising. Moreover, due to the Radon transform, sinogram frequencies exhibit directional coupling, and disrupting specific bands can lead to global artifacts such as distortion or ringing in the final reconstruction. To address this, we propose frequency-adaptive noise scheduling (FANS), which adjusts the spectral distribution of noise over the course of diffusion. Specifically, FANS preserves low-frequency components during early denoising steps to stabilize global structure recovery, and gradually shifts focus to high-frequency perturbations as finer details are restored.

The process begins by generating standard Gaussian noise $\epsilon_t \sim \mathcal{N}(0, I)$, followed by a two-dimensional Fourier transform to obtain its frequency representation $\epsilon_t^{\text{freq}} = \mathcal{F}(\epsilon_t)$. We then apply a time-dependent weighting function $n_t(u,v)$ to modulate the spectral content:
\begin{equation}
    n_t(u, v) = 1 + \left(1 - \frac{t}{T}\right) \cdot \text{dist}(u,v) + \frac{t}{T} \cdot (1 - \text{dist}(u,v)),
\end{equation}
where $\text{dist}(u,v) = \frac{\sqrt{u^2 + v^2}}{d_{\max}}$ represents the normalized distance to the low-frequency origin, and $d_{\max} = \sqrt{(H/2)^2 + (W/2)^2}$ denotes the maximum frequency magnitude. $T$ is the total number of diffusion steps, and $t$ denotes the current step.

The weighted noise is computed via element-wise multiplication and then transformed back to the spatial domain:
\begin{equation}
    \epsilon_t^{\text{adapted}} = \mathcal{F}^{-1}(n_t \cdot \epsilon_t^{\text{freq}}).
\end{equation}

Although FANS improves the spectral targeting of noise, it may still disturb global consistency, particularly in terms of total attenuation along projection paths. To mitigate this, we introduce a soft absorption constraint that penalizes deviations in row-wise intensity while avoiding rigid enforcement that could hurt generalization. While global zero-mean noise is sufficient for RGB images, sinograms require row-wise consistency due to the physical meaning of each projection. Therefore, we correct each row’s mean to reduce bias in total attenuation. Specifically, we compute the row-wise sum of the adapted noise and apply a normalization term. Let $\text{N}[i] = \sum_{j} \epsilon_t^{\text{adapted}}[i, j]$ and $W$ is the width of the sinogram. The final correction is given by:
\begin{equation}
    \epsilon_t^{\text{final}} = \epsilon_t^{\text{adapted}} - \lambda \frac{\text{N}}{W},
\end{equation}
where the subtraction is applied row-wise: each element in row $i$ is reduced by $\frac{\text{N}_i}{W}$. $\lambda = 1$ enforces full projection consistency, $\lambda = 0$ disables correction, and values between offer a balance between physical plausibility and generalization.

The final adjusted noise is injected into the denoising process using the standard diffusion update:
\begin{equation}
    x_t = \sqrt{1 - \beta_t} x_{t-1} + \sqrt{\beta_t} \epsilon_t^{\text{final}}.
\end{equation}

\section{Evaluation}
\label{sec:evaluation}

\begin{figure*}[!ht]
    \centering
    \begin{subfigure}{0.13\textwidth}
        \centering
        \includegraphics[width=\textwidth]{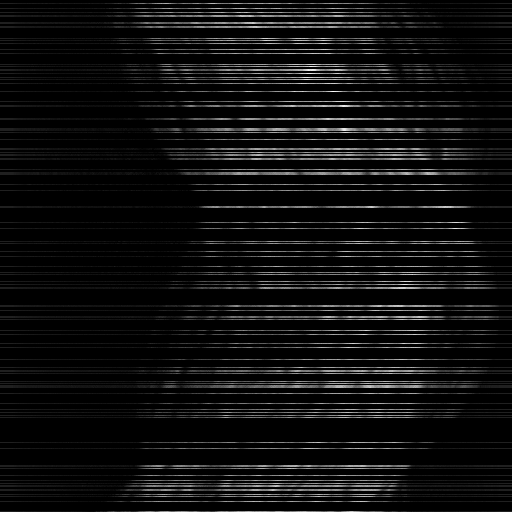}
    \end{subfigure}
    \begin{subfigure}{0.13\textwidth}
        \centering
        \includegraphics[width=\textwidth]{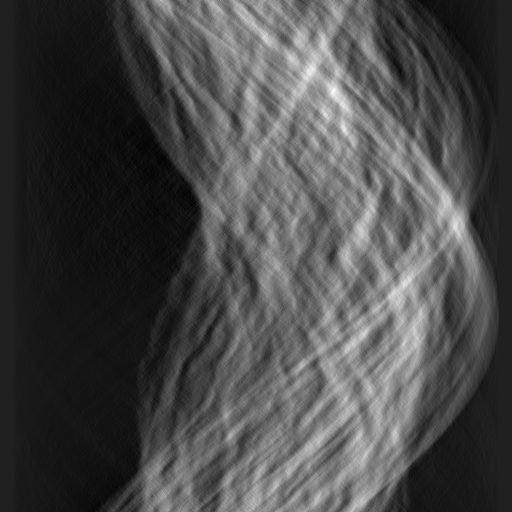}
    \end{subfigure}
    \begin{subfigure}{0.13\textwidth}
        \centering
        \includegraphics[width=\textwidth]{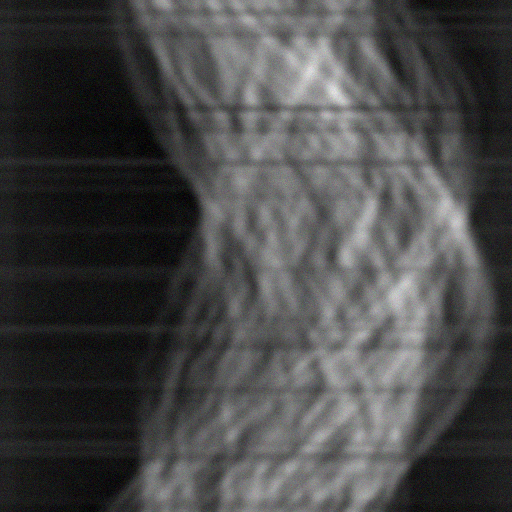}
    \end{subfigure}
    \begin{subfigure}{0.13\textwidth}
        \centering
        \includegraphics[width=\textwidth]{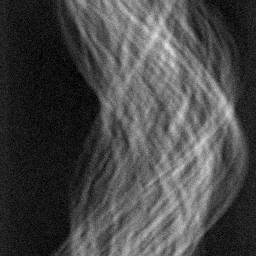}
    \end{subfigure}
    \begin{subfigure}{0.13\textwidth}
        \centering
        \includegraphics[width=\textwidth]{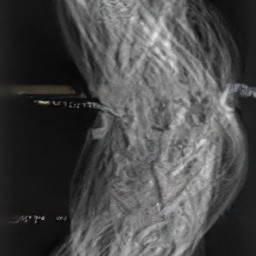}
    \end{subfigure}
    \begin{subfigure}{0.13\textwidth}
        \centering
        \includegraphics[width=\textwidth]{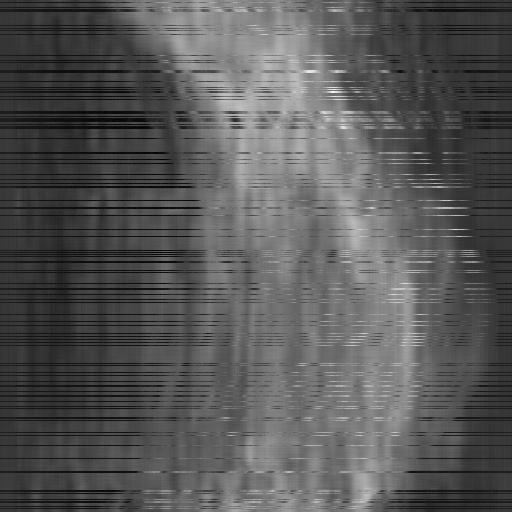}
    \end{subfigure}
    \begin{subfigure}{0.13\textwidth}
        \centering
        \includegraphics[width=\textwidth]{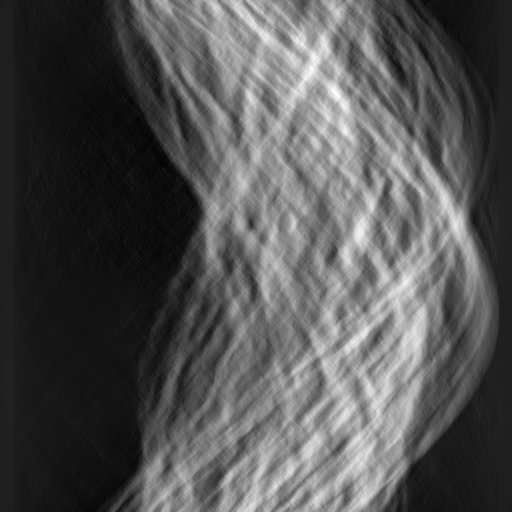}
    \end{subfigure}

    \begin{subfigure}{0.13\textwidth}
        \centering
        \includegraphics[width=\textwidth]{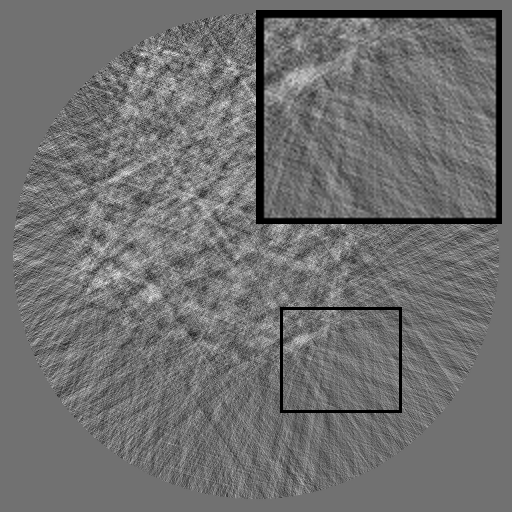}
    \end{subfigure}
    \begin{subfigure}{0.13\textwidth}
        \centering
        \includegraphics[width=\textwidth]{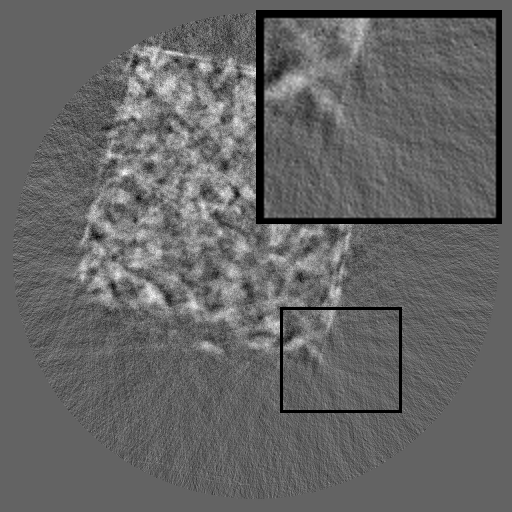}
    \end{subfigure}
    \begin{subfigure}{0.13\textwidth}
        \centering
        \includegraphics[width=\textwidth]{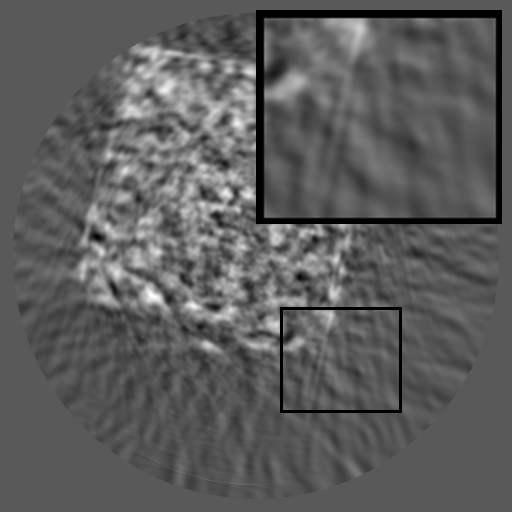}
    \end{subfigure}
    \begin{subfigure}{0.13\textwidth}
        \centering
        \includegraphics[width=\textwidth]{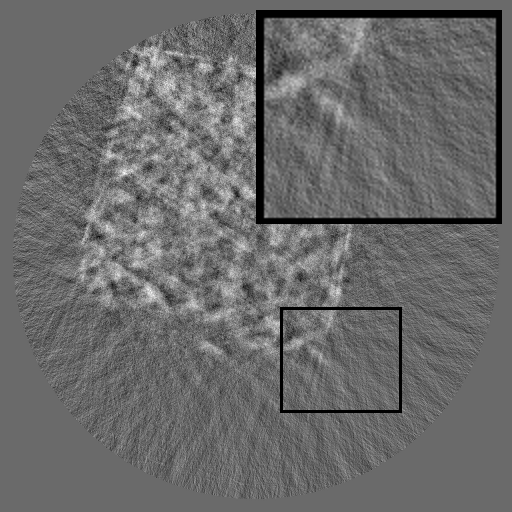}
    \end{subfigure}
    \begin{subfigure}{0.13\textwidth}
        \centering
        \includegraphics[width=\textwidth]{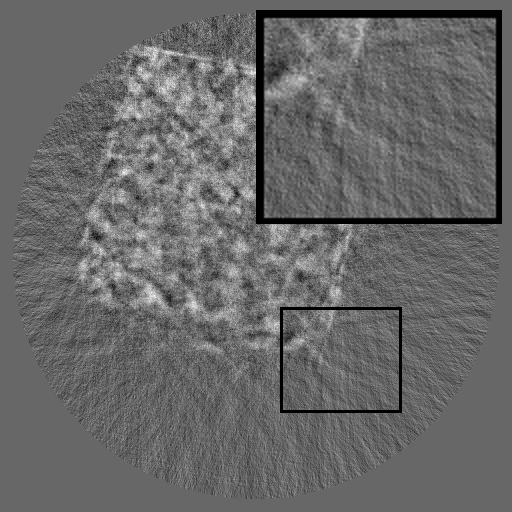}
    \end{subfigure}
    \begin{subfigure}{0.13\textwidth}
        \centering
        \includegraphics[width=\textwidth]{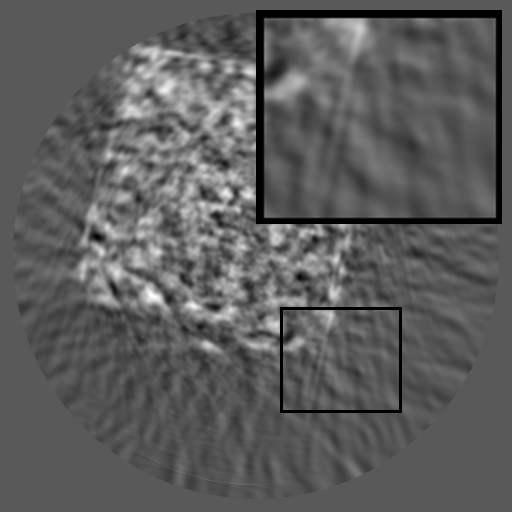}
    \end{subfigure}
    \begin{subfigure}{0.13\textwidth}
        \centering
        \includegraphics[width=\textwidth]{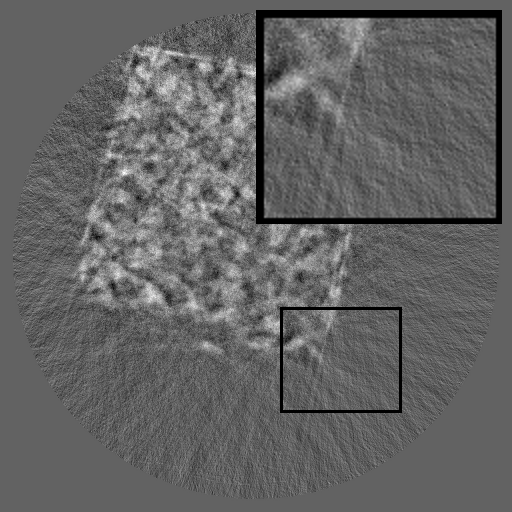}
    \end{subfigure}

    \begin{subfigure}{0.13\textwidth}
        \centering
        \includegraphics[width=\textwidth]{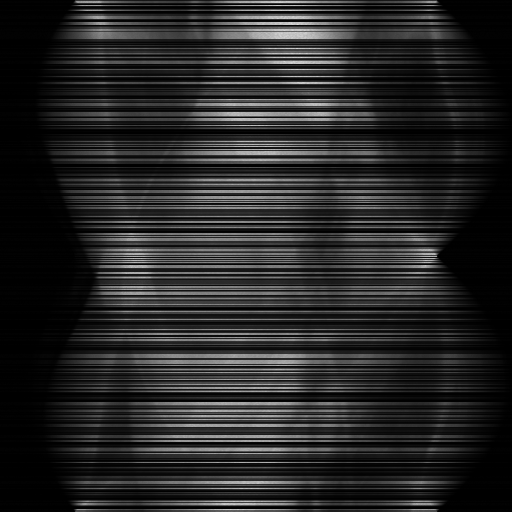}
    \end{subfigure}
    \begin{subfigure}{0.13\textwidth}
        \centering
        \includegraphics[width=\textwidth]{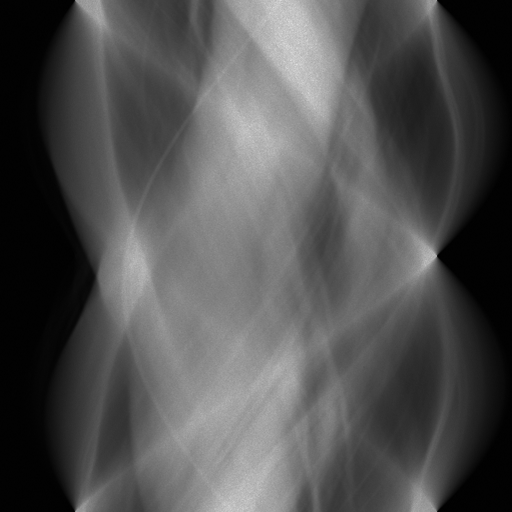}
    \end{subfigure}
    \begin{subfigure}{0.13\textwidth}
        \centering
        \includegraphics[width=\textwidth]{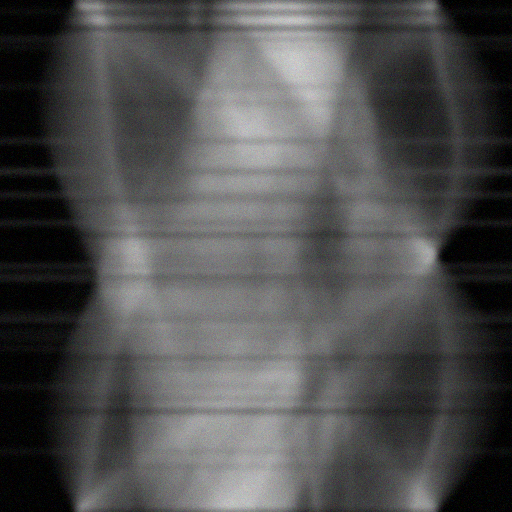}
    \end{subfigure}
    \begin{subfigure}{0.13\textwidth}
        \centering
        \includegraphics[width=\textwidth]{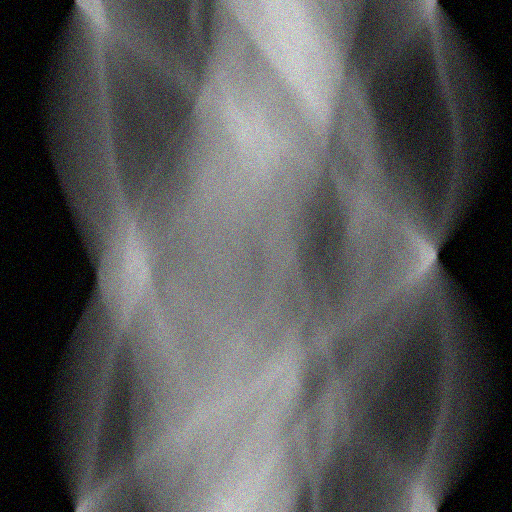}
    \end{subfigure}
    \begin{subfigure}{0.13\textwidth}
        \centering
        \includegraphics[width=\textwidth]{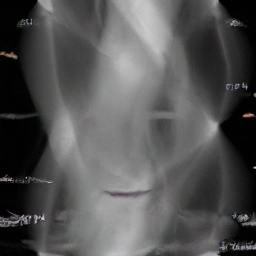}
    \end{subfigure}
    \begin{subfigure}{0.13\textwidth}
        \centering
        \includegraphics[width=\textwidth]{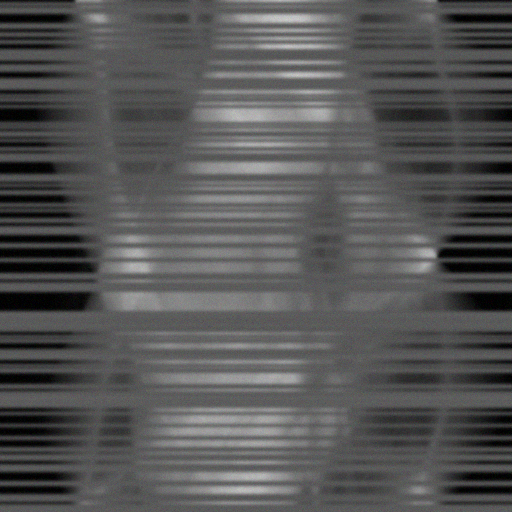}
    \end{subfigure}
    \begin{subfigure}{0.13\textwidth}
        \centering
        \includegraphics[width=\textwidth]{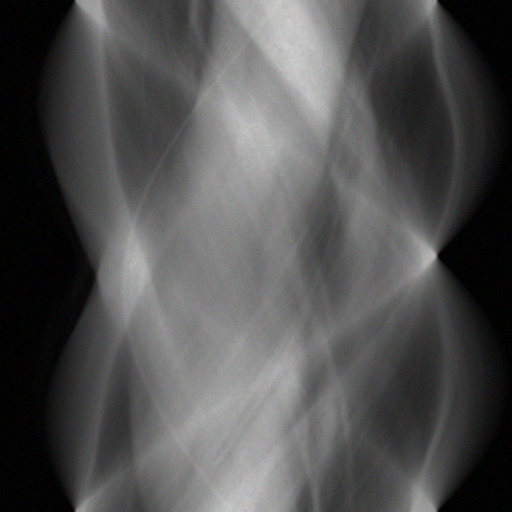}
    \end{subfigure}

    \begin{subfigure}{0.13\textwidth}
        \centering
        \includegraphics[width=\textwidth]{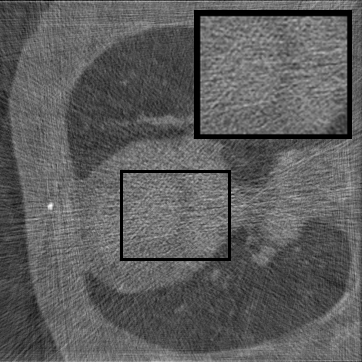}
        \captionsetup{labelformat=empty}
        \caption{Masked}
    \end{subfigure}
    \begin{subfigure}{0.13\textwidth}
        \centering
        \includegraphics[width=\textwidth]{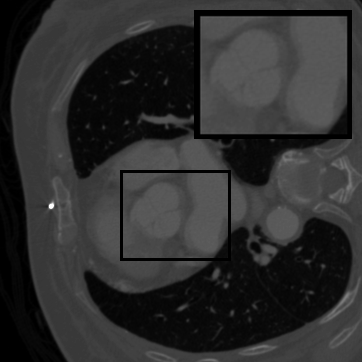}
        \captionsetup{labelformat=empty}
        \caption{Ground Truth}
    \end{subfigure}
    \begin{subfigure}{0.13\textwidth}
        \centering
        \includegraphics[width=\textwidth]{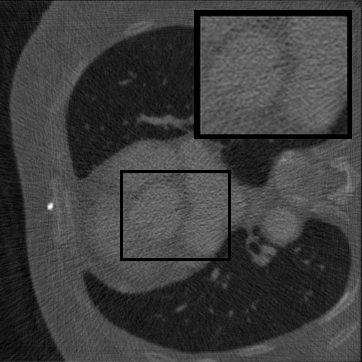}
        \captionsetup{labelformat=empty}
        \caption{LaMa}
    \end{subfigure}
    \begin{subfigure}{0.13\textwidth}
        \centering
        \includegraphics[width=\textwidth]{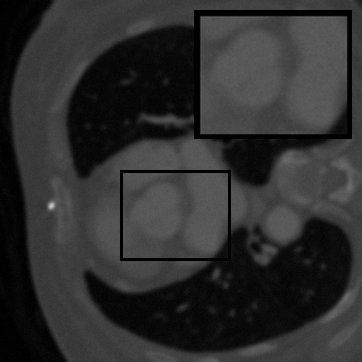}
        \captionsetup{labelformat=empty}
        \caption{RePaint}
    \end{subfigure}
    \begin{subfigure}{0.13\textwidth}
        \centering
        \includegraphics[width=\textwidth]{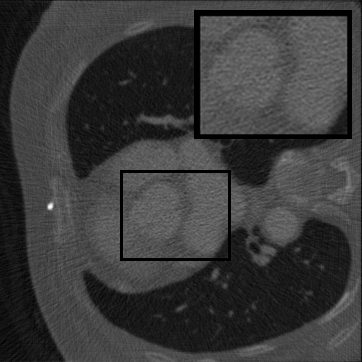}
        \captionsetup{labelformat=empty}
        \caption{CoPaint}
    \end{subfigure}
    \begin{subfigure}{0.13\textwidth}
        \centering
        \includegraphics[width=\textwidth]{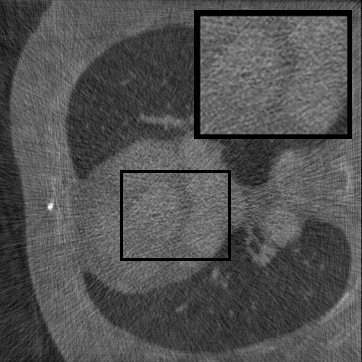}
        \captionsetup{labelformat=empty}
        \caption{SinoTx}
    \end{subfigure}
    \begin{subfigure}{0.13\textwidth}
        \centering
        \includegraphics[width=\textwidth]{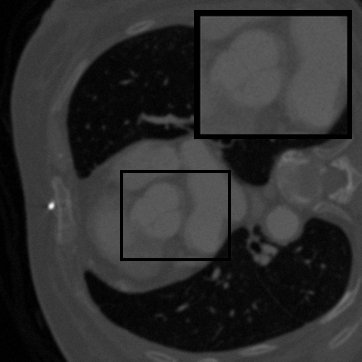}
        \captionsetup{labelformat=empty}
        \caption{FCDM (Ours)}
    \end{subfigure}
    
    \caption{Visual comparison on {\tt TomoBank} and {\tt LoDoPaB} under random masking (ratio = 0.8). Rows 1 and 3 show the inpainted sinograms, while Rows 2 and 4 present the corresponding FBP~\cite{ramachandran1971three}-reconstructed images.}
    \label{fig:comparison}
\end{figure*}

The evaluation has three main objectives: (1) demonstrating that~\modelname~outperforms all baselines on complex real-world datasets (\ref{sec:overall-performance}); (2) conducting ablation studies to validate the effectiveness of our four key designs, including BFDC, physics-guided loss function, FEME, and FANS (\ref{sec:ablation}).

\subsection{Experimental Setup}

\subsubsection{Implementation Details}

We train our framework using the Polaris supercomputer at the Argonne Leadership Computing Facility (ALCF). All training is conducted on four A100 GPUs within one compute node using PyTorch 2.1.0 and CUDA 12.2. We adopt a two-stage training strategy. In the first stage, the encoder–decoder is initialized with pretrained weights from the official latent diffusion models~\cite{rombach2022high} and fine-tuned on our dataset. In the second stage, the autoencoder is frozen and the diffusion module is trained with the same dataset. Each stage is trained for 100 epochs with a batch size of 32 using the AdamW optimizer. The learning rate follows a linear scaling rule with a base value of $2\times10^{-5}$. During training, the mask ratio is randomly selected within [0.1, 0.9] to simulate sparse-view conditions with different sparsity levels. For each training sample, a subset of projection angles is randomly removed, and the corresponding sinogram rows are masked to represent missing views. This random angular masking encourages the framework to learn view-invariant completion and improves generalization across different sparsity ratios.

For BFDC, $\alpha_h$ and $\alpha_w$ are adjusted to 0.6, and 0.4, respectively, as sinograms exhibit smoother variations along the projection angle direction ($W$) but more complex frequency distributions along the detector axis ($H$) \cite{li2020anisotropic}. The loss function is a weighted combination of several components with the following coefficient: pixel loss $\lambda_{pixel} = 1.0$, adversarial loss $\lambda_{adv} = 0.01$, absorption consistency loss $\lambda_{absorp} = 0.1$, and frequency domain loss $\lambda_{freq} = 0.1$. For FEME, we set the Fourier encoding frequencies as $f_i = i$ with $N = 4$. This provides sufficient angular resolution while avoiding redundancy or overfitting in the mask embedding. For FANS, the soft absorption constraint is controlled by $\lambda = 0.5$, allowing limited variations in total absorption to improve generalization.

\subsubsection{Experimental setup}

\begin{table}[t]
    \centering
    \begin{tabular}{ccc}
        \hline
        Method & SSIM & PSNR \\
        \hline
        Ours & \textbf{0.934} & \textbf{30.7} \\
        2D Freq Conv & 0.912 & 28.9 \\
        W/o $W$ Conv & 0.906 & 28.6 \\
        W/o $H$ Conv & 0.901 & 28.1 \\
        Spatial Conv & 0.888 & 27.7 \\
        W/o Freq Conv & 0.880 & 27.4 \\
        \hline
    \end{tabular}
    \caption{Ablation results for frequency-domain convolutions on the {\tt TomoBank} dataset under random masks (ratio = 0.8). Removing frequency-domain convolutions leads to a sharp decline in performance, indicating their necessity for disentangling spectral overlap. The variant using only $H$-axis convolutions performs slightly better than that using $W$-axis convolutions, but the full BFDC achieves the best overall results.}
    \label{tab:ablation-freq-conv}
\end{table}

\noindent{\bf Datasets.}
We evaluate our method on two real-world datasets, {\tt TomoBank} and {\tt LoDoPaB}, each containing 100k samples. Both datasets are split into 80\% training, 10\% validation, and 10\% testing, with all samples at a resolution of $512 \times 512$. {\tt TomoBank} \citep{de2018tomobank} consists of real sinograms acquired from synchrotron radiation CT experiments on diverse materials and objects, primarily at the Advanced Photon Source (APS) at ANL, and includes dynamic \citep{mohan2015timbir} and in situ \citep{pelt2013fast} measurements. We follow the official TomoPy \citep{gursoy2014tomopy} preprocessing pipeline, including ring artifact removal, rotation center alignment, and intensity normalization to [0,1], with projection counts, angular sampling, and rotation centers configured according to the provided metadata. {\tt LoDoPaB} \citep{leuschner2021lodopab} is derived from the LIDC-IDRI lung CT dataset \citep{armato2011lung} and provides numerically generated sinograms paired with real patient CT images, using a fixed parallel-beam geometry (1,000 projection angles, 513 detector bins). Unlike TomoBank, which contains physically measured sinograms, LoDoPaB offers simulated data with official splits provided via the DIVAL benchmark library \citep{DIVAL}.

\noindent{\bf Metrics.}
We evaluate two masking schemes—random and periodic—with mask ratios of 0.4, 0.6, and 0.8 to simulate varying degrees of sparse-view sampling. The random mask models irregular acquisition patterns, while the periodic mask represents fixed-step angular undersampling in hardware-limited CT systems. We evaluate inpainting quality using the Structural Similarity Index Measure (SSIM) and Peak Signal-to-Noise Ratio (PSNR). SSIM measures structural preservation by comparing local luminance, contrast, and texture, while PSNR assesses pixel-wise reconstruction fidelity. To assess reconstruction fidelity, we further apply filtered back-projection (FBP) \cite{ramachandran1971three} to the inpainted sinograms and compute the same metrics on the reconstructed images, enabling consistent evaluation at both the projection and reconstruction levels.

\noindent{\bf Baselines.}
We compare~\modelname~with ten representative baselines, covering three categories. (1) Diffusion-based inpainting methods: RePaint \cite{lugmayr2022repaint}, CoPaint \cite{zhang2023coherent}, and StrDiffusion \cite{liu2024structure}. (2) Sinogram-specific learning methods: SinoTx \cite{jiaze2025sinotx}, UsiNet \cite{yao2024no}, and CoIL \cite{9606601}. (3) non-diffusion inpainting methods: MISF \cite{li2022misf}, LaMa \cite{suvorov2022resolution}, Hourglass \cite{deng2022hourglass}, and CMT \cite{ko2023continuously}. This categorization allows us to compare~\modelname~both against diffusion-based generative methods and traditional or task-specific baselines.

\subsection{Accuracy Comparisons with Baselines}
\label{sec:overall-performance}

This section evaluates~\modelname's performance by comparing it against 10 baselines on two real-world datasets. We focus on evaluating the SSIM and PSNR of the sinograms and also images reconstructed by FBP \cite{ramachandran1971three}.

\begin{table}
    \centering
    \setlength{\tabcolsep}{5pt}
    \begin{tabular}{ccc}
        \hline
        Method & SSIM & PSNR \\
        \hline
        Ours & \textbf{0.934} & \textbf{30.7} \\
        W/o $\mathcal{L}_{adv}$ & 0.918 & 28.9 \\
        W/o $\mathcal{L}_{freq}$ & 0.913 & 28.3 \\
        W/o $\mathcal{L}_{absorp\_sum}$ & 0.909 & 27.8 \\
        \hline
    \end{tabular}
    \caption{Ablation results on tailored loss functions on the {\tt TomoBank} dataset under random masks (ratio = 0.8). Excluding either the absorption consistency loss ($L_{absorp}$) or the frequency-domain loss ($L_{freq}$) results in clear degradation, verifying that both losses jointly contribute to the inpainting process.}
    \label{tab:ablation-loss}
\end{table}

Table~\ref{tab:comparison} summarizes the quantitative performance of~\modelname~and all baselines on the {\tt TomoBank} and {\tt LoDoPaB} datasets under both random and periodic masks. Figure~\ref{fig:comparison} presents qualitative comparisons at a mask ratio of 0.8, showing both the inpainted sinograms and the reconstructed images obtained using FBP \cite{ramachandran1971three}. Across all settings,~\modelname~consistently achieves the highest SSIM and PSNR values, demonstrating its strong capability in recovering structural details and suppressing noise-induced artifacts. Visual results in Figure~\ref{fig:comparison} further confirm that~\modelname~produces smoother and more consistent sinograms, while other methods often exhibit angular streaking or local inconsistencies.

\begin{table}
    \centering
    \begin{tabular}{ccc}
        \hline
        Method & SSIM & PSNR \\
        \hline
        Ours & \textbf{0.934} & \textbf{30.7} \\
        W/o FEME & 0.913 & 28.6 \\
        W/o FANS & 0.908 & 27.6 \\
        \hline
    \end{tabular}
    \caption{Ablation results for FEME and FANS on the {\tt TomoBank} dataset under random masks (ratio = 0.8). Removing either module leads to a consistent drop in SSIM and PSNR, indicating that both components contribute to improved sinogram restoration quality. The combination of FEME and FANS achieves the best overall results across all evaluation metrics.}
    \label{tab:ablation-feme-fans}
\end{table}

\subsection{Ablation Studies}
\label{sec:ablation}

To analyze the effectiveness of individual components, we conduct ablation experiments on the {\tt TomoBank} dataset under random masks with a ratio of 0.8 by comparing the sinogram inpainting results.

\noindent{\bf Impact of Bidirectional Frequency Domain Convolutions.}
We first investigate the influence of BFDC by comparing different convolutional configurations. Variants include spatial-only convolutions, full 2D frequency convolutions, and single-direction frequency convolutions along either the detector ($H$) or projection-angle ($W$) dimensions. As shown in Table~\ref{tab:ablation-freq-conv}, removing frequency-domain convolutions causes a notable performance drop, confirming that spatial convolutions alone cannot effectively disentangle overlapping features in sinograms. Among the frequency-based variants, using convolutions only along the detector axis ($H$) slightly outperforms those along the angular axis ($W$), reflecting that structural variations are generally more complex along the detector dimension. The full BFDC configuration, which jointly models both directions, achieves the best results in both SSIM and PSNR.

\noindent{\bf Contribution of Tailored Loss Functions.}
We next examine the contribution of the physics-guided and frequency-domain loss terms by selectively removing them. Table~\ref{tab:ablation-loss} shows that excluding either the absorption consistency loss ($L_{absorp}$) or the frequency-domain regularization loss ($L_{freq}$) significantly degrades inpainting quality. The combination of both terms yields the most stable and physically consistent reconstructions.

\noindent{\bf Impact of Fourier-Enhanced Mask Embedding and Frequency-Adaptive Noise Scheduling.}
Finally, we evaluate the contribution of the diffusion-specific components, FEME and FANS. As shown in Table~\ref{tab:ablation-feme-fans}, removing either component leads to consistent drops in SSIM and PSNR, confirming their effectiveness in improving sinogram restoration quality. Using standard binary masks instead of FEME results in weaker performance, suggesting that incorporating angular information provides additional contextual cues for inpainting. Similarly, replacing FANS with standard Gaussian noise also degrades performance, indicating that a frequency-aware noise schedule better matches the sinogram characteristics.

\section{Conclusion}

This work introduces~\modelname, a new diffusion-based sinogram inpainting framework that integrates bidirectional frequency domain convolutions, physics-guided loss, Fourier-enhanced mask embedding, and frequency-adaptive noise scheduling. Evaluations demonstrate that~\modelname~closely approximates ground truth. The ablation studies then further validate the effectiveness of each component.

\bibliography{example_paper}
\bibliographystyle{icml2026}

\newpage
\appendix
\onecolumn

\section{Design Rationales of Key Components}

\modelname~is composed of several interacting components, each designed to address a specific aspect of the sinogram inpainting challenge. While the main paper has already discussed the motivations, this section provides further clarification on the reasoning behind their structural choices or usage contexts.

\subsection{Bidirectional Frequency-Domain Convolution in Latent Space}

A central design choice in~\modelname~is the placement of the bidirectional frequency-domain convolution (BFDC): whether it should operate directly on the input sinogram or on the latent representation produced by the encoder. While applying BFDC at the input level may seem attractive due to access to full-resolution data, we argue that performing it in the latent space is both intuitively and functionally more appropriate for our task.

Sinogram data, particularly from real-world sources, contain a significant amount of noise and physically irrelevant high-frequency components. Applying frequency-domain filtering directly to this raw input risks amplifying these artifacts and misinterpreting noise as structural signal. In contrast, the encoder in~\modelname~is trained not merely to reduce spatial dimensions, but to abstract physically meaningful features while suppressing noise and redundancy. The resulting latent space provides a cleaner, semantically richer representation of the underlying projection structure.

This design choice aligns with established practices in modern generative models and vision transformers, where attention or frequency-based operations are often applied at the feature level rather than on raw-resolution input \cite{suvorov2022resolution, chi2020fast}, allowing better generalization and noise robustness.

\subsubsection{Theoretical Complexity Analysis}

We briefly compare the theoretical complexity of applying BFDC at different stages of the model pipeline. In the first setting, BFDC is applied directly on the input sinogram of size $C \times H \times W$, followed by an encoder. In the second setting, the encoder first maps the input into a latent feature space of size $C_L \times H_L \times W_L$, after which BFDC is applied.

In the input-level setting, the dominant operations include the forward and inverse Fourier transforms and the element-wise multiplication in the frequency domain, followed by spatial convolution and the standard encoder. The corresponding total complexity can be approximated as:
\begin{equation}
    \mathcal{O}\left(C \cdot HW \cdot \log HW + C^2 k^2 HW\right)
\end{equation}
Here, $k$ denotes the spatial kernel size used in the convolution operations.

In the latent-level setting, the encoder is applied first, and BFDC is performed on its output. The overall complexity then becomes:
\begin{equation}
\mathcal{O}\left(C^2 k^2 HW + C_L \cdot \tfrac{HW}{r^2} \cdot \log \tfrac{HW}{r^2} + C_L^2 k^2 \tfrac{HW}{r^2} \right)
\end{equation}
Here, $r$ denotes the effective downsampling ratio of the encoder, and we assume that both versions use the same encoder architecture. While the encoder cost is present in both settings, the frequency-related operations are significantly reduced in the latent case due to the smaller spatial resolution and potentially lower channel dimensionality. These expressions suggest that placing BFDC in the latent space is computationally advantageous, particularly when working with high-resolution inputs.

\subsubsection{AutoEncoder-Level Performance Analysis}

To further assess the practical impact of BFDC placement, we compare the runtime efficiency and inpainting quality of two AutoEncoder variants: one applying BFDC directly on the input sinogram (input-level), and the other applying it on the encoded latent features (latent-level). This analysis isolates the AutoEncoder module from the full model, excluding any diffusion-based refinement or masking strategies, and focuses solely on the encoder-decoder backbone.

All AutoEncoder-level experiments are conducted in inference mode using the same implementation setup as in the main paper. We apply standard performance optimizations, including {\tt torch.compile()} for graph-mode execution, {\tt torch.autocast()} for mixed-precision (fp16) inference, and {\tt cudnn.benchmark = True} for automatic kernel selection. All evaluations are performed with batch size 1 on the {\tt TomoBank} dataset.

\begin{table}[ht]
    \centering
    \begin{tabular}{lccccc}
        \toprule
        Variant & SSIM & PSNR & Mem & Time & FLOPs\\
        \midrule
        @Input  & 0.948 & 32.5 & 3.7 & 0.37 & 18.7\\
        @Latent & 0.955 & 33.1 & 3.7 & 0.36 & 18.5\\
        W/o & 0.904 & 29.3 & 3.6 & 0.33 & 18.4\\
        \bottomrule
    \end{tabular}
    \caption{AutoEncoder-only comparison between input-level, latent-level, and no-BFDC variants. Results are measured on the {\tt TomoBank} dataset with batch size 1. Memory, time, and FLOPs denote maximum allocated GPU memory (GB), per-sinogram inference time (s), and theoretical computation cost (G), respectively.}
    \label{tab:autoencoder-performance}
\end{table}

We additionally include a variant where BFDC is fully removed, yielding a purely spatial AutoEncoder baseline. As shown in Table~\ref{tab:autoencoder-performance}, the latent-level variant attains the highest full-sinogram prediction quality, exceeding both the input-level variant and the spatial baseline. In terms of efficiency, input-level BFDC incurs slightly higher computational cost compared with latent-level placement, which operates on reduced-resolution features and therefore runs marginally faster. The spatial baseline is the cheapest but exhibits a clear drop in accuracy. These observations indicate that performing BFDC in the latent space offers the most favorable accuracy–efficiency trade-off, balancing frequency-domain expressiveness with reduced computational overhead.

\subsection{Absorption Consistency via Inverse Radon}

Our proposed physics-guided loss contains a total absorption consistency term, referred to as $L_{absorp}$, which enforces the principle that the cumulative attenuation along each X-ray path should remain stable across the projection domain. This principle reflects a fundamental property of X-ray imaging: the amount of total energy absorbed by a scanned object should be preserved globally.

To implement this constraint, we apply an inverse Radon transform to the predicted sinogram, producing an approximate attenuation map. Although the formulation uses analytic operations rather than full physical modeling, it approximates a domain-consistent property rooted in imaging physics. The inverse Radon transform used here is not intended as an accurate simulation of material attenuation, but rather as a differentiable surrogate that allows the network to reason about global attenuation structure.

This design reflects a broader philosophy of physics-guided learning: rather than modeling the full measurement process, we incorporate physics-motivated constraints into the loss function, using approximate but structured operations to inform optimization.

\section{Hyperparameter Design and Sensitivity}

\subsection{Axis Weighting in BFDC}

Our BFDC module applies axis-specific weighting to frequency components along the projection angle ($W$) and detector ($H$) dimensions. This is controlled by two hyperparameters: $a_w$, the weight for angular (row-wise) convolution, and $a_h$, the weight for detector (column-wise) convolution. In our main experiments, we set $\alpha_h = 0.6$ and $\alpha_w = 0.4$, based on the observation that sinograms typically exhibit smoother variation along the angular direction but more complex high-frequency structure along the detector axis.

To assess the influence of directional weighting in BFDC, we perform an ablation in the AutoEncoder-only full-sinogram prediction setting, where the network is trained to map each sinogram to itself without masking. All evaluations are conducted on the {\tt TomoBank} dataset. We compare the default configuration used in the main paper ($\alpha_h = 0.6,\ \alpha_w = 0.4$) with two representative alternatives: a balanced isotropic variant ($\alpha_h = \alpha_w = 0.5$) and a swapped angular-dominant variant ($\alpha_h = 0.4,\ \alpha_w = 0.6$). All other components follow the same implementation as in the main paper.

\begin{table}[ht]
    \centering
    \begin{tabular}{lcc}
        \toprule
        BFDC Configuration & SSIM & PSNR \\
        \midrule
        Original & \textbf{0.955} & \textbf{33.1} \\
        Isotropic & 0.946 & 32.1 \\
        Angular-dominant  & 0.940 & 31.7 \\
        \bottomrule
    \end{tabular}
    \caption{Effect of BFDC axis weighting in the AutoEncoder-only full-sinogram prediction setting. Results are measured on the {\tt TomoBank} dataset.}
    \label{tab:bfdc-hyperparams}
\end{table}

As shown in Table~\ref{tab:bfdc-hyperparams}, both the isotropic setting and the swapped angular-dominant setting yield a consistent decrease in SSIM and PSNR compared with the default configuration. Treating the two axes identically or reversing their relative importance degrades full-sinogram prediction quality, indicating that the axial imbalance encoded by BFDC is a meaningful bias reflecting the anisotropic structure of sinograms. These findings support our choice of axis-aware frequency weighting.

\subsection{Soft Absorption Weight in FANS}

To encourage projection-level consistency during denoising, we incorporate a soft absorption constraint into FANS, which penalizes deviations in row-wise attenuation. The strength of this constraint is controlled by a hyperparameter $\lambda$, balancing physical regularization and model flexibility. Large $\lambda$ values may over-constrain the denoising trajectory, while very small $\lambda$ values make the constraint ineffective.

To assess the influence of this hyperparameter, we compare three representative configurations—no constraint ($\lambda = 0.0$), the default moderate setting ($\lambda = 0.5$), and a strong constraint ($\lambda = 1.0$). Experiments follow the same implementation setup as in the main paper and are conducted on the {\tt TomoBank} dataset using a random mask with ratio 0.8.

\begin{table}[ht]
    \centering
    \begin{tabular}{lcc}
        \toprule
        $\lambda$ & SSIM & PSNR \\
        \midrule
        0.0 (no constraint) & 0.925 & 29.7 \\
        0.5 (default)       & \textbf{0.934} & \textbf{30.7} \\
        1.0 (hard constraint) & 0.927 & 29.9 \\
        \bottomrule
    \end{tabular}
    \caption{Effect of the soft absorption constraint in FANS. A moderate value of $\lambda$ provides the best balance between enforcing projection-level consistency and maintaining robust inpainting performance.}
    \label{tab:lambda-ablation}
\end{table}

As shown in Table~\ref{tab:lambda-ablation}, the moderate setting ($\lambda = 0.5$) achieves the best inpainting performance, outperforming both the unconstrained variant ($\lambda = 0.0$) and the over-constrained variant ($\lambda = 1.0$). The performance drops observed at the extremes indicate that the absorption constraint provides meaningful guidance without dominating the denoising process. These results demonstrate that a soft, mid-range value of $\lambda$ is sufficient to enforce projection-level consistency while preserving the model’s generalization ability.

\section{Additional Experimental Notes}

In sinogram-based CT, large contiguous missing regions do not correspond to realistic acquisition conditions. Each sinogram row represents a full projection at a specific rotation angle, and removing a continuous block of rows implies that an entire angular range was never measured. Unlike RGB image inpainting—where a missing region may admit many visually plausible completions—CT acquisition is a deterministic physical process: a sinogram encodes quantitative attenuation measurements that must correspond to a single, physically valid object. When a continuous set of projection angles is absent, the resulting completion problem becomes fundamentally ill-posed, yielding no unique or physically meaningful sinogram that can support downstream reconstruction. For this reason, sparse-view CT research consistently focuses on random or periodic view-sampling patterns, and we adopt these standard settings.

To evaluate the computational footprint of~\modelname~during inference, we compare its runtime and peak memory usage with representative diffusion-based inpainting baselines. All methods are evaluated under the same random mask with a mask ratio of 0.8. DDIM sampling with 50 steps is used for all diffusion models. Hardware and software configurations are kept consistent with those used throughout the paper.

\begin{table}[ht]
    \centering
    \begin{tabular}{lcc}
        \toprule
        Methods & Mem & Time \\
        \midrule
        Ours & 5.1 & 0.86 \\
        RePaint & 5.5 & 1.44\\
        StrDiffusion & 5.2 & 0.90 \\
        \bottomrule
    \end{tabular}
    \caption{Inference efficiency comparison with representative baselines. Memory is reported as peak GPU allocation in gigabytes (GB), and time is reported as per-sinogram inference time in seconds (s).}
    \label{tab:system}
\end{table}

As shown in Table~\ref{tab:system},~\modelname~achieves inference efficiency comparable to recent diffusion-based inpainting models. It is substantially faster than RePaint, which relies on iterative resampling, and remains on par with StrDiffusion while using slightly less memory. These results indicate that~\modelname~does not impose disproportionate computational overhead relative to representative diffusion baselines.

\section{Limitations}

While~\modelname~demonstrates strong performance across diverse sinogram inpainting scenarios, it still has several limitations. First, our model assumes that each projection view contains valid information across the full detector span. In extreme cases where entire vertical stripes (i.e., detector regions) are missing across all angles, recovery may fail to produce plausible completions. Second,~\modelname~is trained under the assumption of monotonically ordered and unique projection angles. When the projection angles are shuffled, non-uniformly spaced, or duplicated, performance may degrade due to disrupted angular context. Third, as a generative model,~\modelname~may in rare cases hallucinate plausible but physically incorrect structures, particularly under extreme undersampling or ambiguous contexts. While our physics-guided losses aim to reduce such risks, addressing the broader issue of hallucination or inverse hallucination remains an important direction for future work, especially for safety-critical domains.


\end{document}